\documentclass[preprint,notoc]{JHEP3}
\usepackage{epsfig}

\def\to{\rightarrow}

\def\bi{\begin{itemize}}
\def\ei{\end{itemize}}

\def\tst{\tilde t}
\def\ttau{\tilde \tau}

\def\tw{\widetilde W}
\def\tz{\widetilde Z}
\def\alt{\stackrel{<}{\sim}}
\def\agt{\stackrel{>}{\sim}}

\title{Chi**2 analysis of the minimal supergravity model including WMAP, 
g(mu)-2 and b$\to$ s gamma constraints}

\author{Howard Baer and  Csaba Bal\'azs
\\ Department of Physics, Florida State University\\ 
Tallahassee, FL, USA 32306\\
E-mail: \email{baer@hep.fsu.edu}, \email{balazs@hep.fsu.edu}}

\preprint{\vbox{\hbox{FSU-HEP-030313}}}

\abstract{Recent results from the WMAP measurements of the cosmic
background radiation yield very tight constraints on the relic density
of supersymmetric cold dark matter. We combine the WMAP constraint
with those from the anomalous magnetic moment of the muon and the
$b\to s\gamma$ branching fraction in a $\chi^2$ determination over
the minimal supergravity model (mSUGRA) parameter space. 
The most favored region of mSUGRA
parameter space for almost all $\tan\beta$ values is the 
hyperbolic branch/focus point (HB/FP) region, 
with  moderate to small values of 
superpotential Higgs mass $|\mu |$ and large GUT scale scalar mass $m_0$.
These favored regions of mSUGRA parameter space can be probed by
direct search experiments for supersymmetric dark matter. An 
exception to the HB/FP region can occur at very large $\tan\beta$
with positive $\mu $ values, where wide regions allow resonance 
annihilation of neutralinos in the early universe.
}


\begin{document}

The past decade has witnessed increasingly precise measurements
of the anisotropies of the cosmic microwave background radiation
left over from the Big Bang\cite{cmb}. 
The most recent results come from the
Wilkinson Microwave Anisotropy Probe (WMAP) satellite measurements.
Astonishingly, an analysis of their results pinpoints the age of the
universe to be $13.7\pm 0.2$ Gyrs\cite{wmap}. 
In addition, the geometry of the universe
is flat, consistent with simple inflationary models. The dark energy content
of the universe is found to be about $73\%$, while the matter content is 
about $27\%$. A best fit of WMAP and other data sets to cosmological 
parameters in the $\Lambda CDM$ cosmological model yields a
determination of baryonic matter density $\Omega_b h^2=0.0224\pm0.0009$,
a total matter density of $\Omega_mh^2=0.135^{+0.008}_{-0.009}$, 
and a very low density of hot dark matter (relic neutrinos). From
these values the cold
dark matter (CDM) density of $\Omega_{CDM}h^2=0.1126^{+0.0161}_{-0.0181}$
(at $2\sigma$) can be inferred.
The new WMAP results can thus be used to obtain more severe constraints
on particle physics models that include candidates for cold dark matter,
such as supersymmetric theories\cite{eoss}.

It is well known that the lightest supersymmetric particle (LSP) of many 
supersymmetric models has the necessary attributes to make up the
bulk of cold dark matter in the universe\cite{susydm}.
This holds true especially in supergravity models where supersymmetry 
breaking occurs in a hidden sector of the model\cite{sugra}. 
SUSY breaking is 
communicated to the observable sector via gravitational interactions, 
leading to soft SUSY breaking mass terms which can be of order $\sim 1$ TeV,
so that the hierarchy between the weak scale and any other high scale
such as $M_{GUT}$ or $M_{Pl}$ can be stabilized.
The simplest of these models assumes a flat Kahler metric and a simple form
for the gauge kinetic function at the high scale. Here, motivated by gauge
coupling unification at $M_{GUT}\simeq 2\times 10^{16}$ GeV, we thus 
assume common scalar masses $m_0$, common gaugino masses $m_{1/2}$, 
and common trilinear terms $A_0$ all valid at scale $Q=M_{GUT}$.
Below $M_{GUT}$, the effective theory is assumed to be the minimal 
supersymmetric standard model (MSSM). The weak scale sparticle masses
and couplings are determined by renormalization group running between
$M_{GUT}$ and $M_{weak}$, which leads to radiative electroweak
symmetry breaking (REWSB). The mSUGRA model parameters
\begin{equation}
m_0,\ m_{1/2},\ A_0,\ \tan\beta\ \ {\rm and}\ \ sign(\mu)
\end{equation}
then determine all superparticle and Higgs boson masses and mixings.
Here, $\tan\beta$ is as usual the ratio of Higgs field vevs. 
We use the program ISAJET v7.64p\footnote{Isajet 7.64p is
Isajet 7.64 modified to gain access to low $\mu$ sparticle mass 
solutions.} for our sparticle mass 
calculations\cite{isajet}.

Once the sparticle masses and mixings are determined, a variety of
observable quantities can be calculated. In this letter, we focus
especially on the neutralino relic density $\Omega_{\tz_1}h^2$. 
The relic abundance of neutralinos can be calculated by solving the Boltzmann 
equation as formulated for a Friedmann-Robertson-Walker universe.
Central to this calculation is the evaluation of the neutralino
annihilation and co-annihilation cross sections, which must then be 
convoluted with the thermal distribution of neutralinos 
(and possibly other co-annihilating particles) present in the 
early universe. We adopt the calculation of Ref.~\cite{bbb}, wherein
all relevant annihilation and co-annihilation reactions are included
along with relativistic thermal averaging\cite{gg}
(see also Ref. \cite{edsjo} for a recent relic density calculation). 
Four regions of mSUGRA model
parameter space emerge where the CDM relic density is consistent with
measurements. These include A.) a bulk region at low $m_0$ and low $m_{1/2}$,
where neutralino annihilation occurs mainly via $t$-channel
slepton exchange, B.) the stau co-annihilation region where
$\tz_1-\ttau_1$ and $\ttau_1-\bar{\ttau}_1$ annihilations 
contribute\cite{ellis}, 
C.) a region where $2m_{\tz_1}\sim m_{A,H}$, where neutralinos can annihilation
via $s$-channel pseudoscalar ($A$) and heavy scalar ($H$) Higgs 
bosons\cite{Apole},
and D.) the 
region at large $m_0$ where $|\mu |$ becomes
small\cite{bcpt} (known as the hyperbolic branch/focus point (HB/FP) 
region\cite{ccn,focus}), 
and the growing higgsino component of $\tz_1$ allows for
efficient neutralino annihilation and 
co-annihilation\cite{fmw}.\footnote{In this paper, we consider mSUGRA 
solutions with $A_0=0$ only. For particular $A_0$ values, the value of 
$m_{\tst_1}$ can be dialed to near degeneracy with $m_{\tz_1}$, so that 
a fifth region of stop-neutralino co-annihilation occurs\cite{stop}.}

Another important constraint on the mSUGRA model comes from
comparison of the predicted rate for $b\to s\gamma$ decay against 
experimental measurements. Here, we adopt the branching fraction
$BF(b\to s\gamma )=(3.25\pm 0.54)\times 10^{-4}$ for our analysis, and 
use the theoretical evaluation given in Ref. \cite{bsg,sugcon}.
Generally, the value of $BF(b\to s\gamma )$ calculated in the mSUGRA model
differs most from the SM prediction in the region of low $m_0$ and 
$m_{1/2}$, where sparticle masses are light, and SUSY loop contributions 
can be large.

The recently improved measurement of the muon anomalous magnetic moment 
$a_\mu =(g-2)_\mu$ also provides an important constraint on supersymmetric
models\cite{e821}. A recent determination of the deviation between the
measured value of $a_\mu$ and the SM prediction has been made by Narison,
including additional scalar meson loops\cite{narison}. His determination
using $e^+e^-\to hadrons$ data to evaluate hadronic vacuum 
polarization contributions yields 
$\Delta a_\mu =(24.1\pm 14.0)\times 10^{-10}$, which we adopt for this
analysis\cite{gm2}. An alternative determination using $\tau$-decay data
may include additional systematic uncertainties, and is usually 
considered less reliable.

Finally, we include in our determination of allowed parameter
space direct superparticle and Higgs boson search results from the 
LEP2 experiments. The most important of these is that
$m_{\tw_1}>103.5$ GeV on the lightest chargino\cite{lep2_w1}, 
and $m_h>114.1$ GeV
when the lightest Higgs boson is SM-like\cite{lep2_h}. In regions where $m_A$
is small, this bound may be considerably reduced to 
$m_h\agt 90$ GeV, depending as well on the value of $\tan\beta$.

In this analysis, we compute a $\chi^2$ value constructed
from the mSUGRA model calculated values of $\Omega_{\tz_1}h^2$, 
$BF(b\to s\gamma )$ and $a_\mu^{SUSY}$, along with the above mentioned
central values and error bars\cite{deboer}. If the relic density
$\Omega_{\tz_1}h^2$ falls below the WMAP central value, then we do not include
it in our $\chi^2$ determination since other forms of CDM may be present.
Thus, at each point in model parameter space, we
determine the value of $\chi^2$, and represent the value $\log (\chi^2 /3)$
by various colors in the $m_0\ vs.\ m_{1/2}$ plane for different values
of $\tan\beta$ and $sign (\mu )$. 
The green regions generally correspond to a $\chi^2/dof$ value less than 
$4/3$,
while yellow regions have $4/3 \alt \chi^2/dof \alt 25/3$. The yellow regions 
shade into red for larger $\chi^2/dof$ values.
We adopt $A_0=0$ throughout our analysis.
In general, our conclusions do not change qualitatively upon variation
of $A_0$, unless extreme values of the parameter are adopted.

\begin{figure}
\epsfig{file=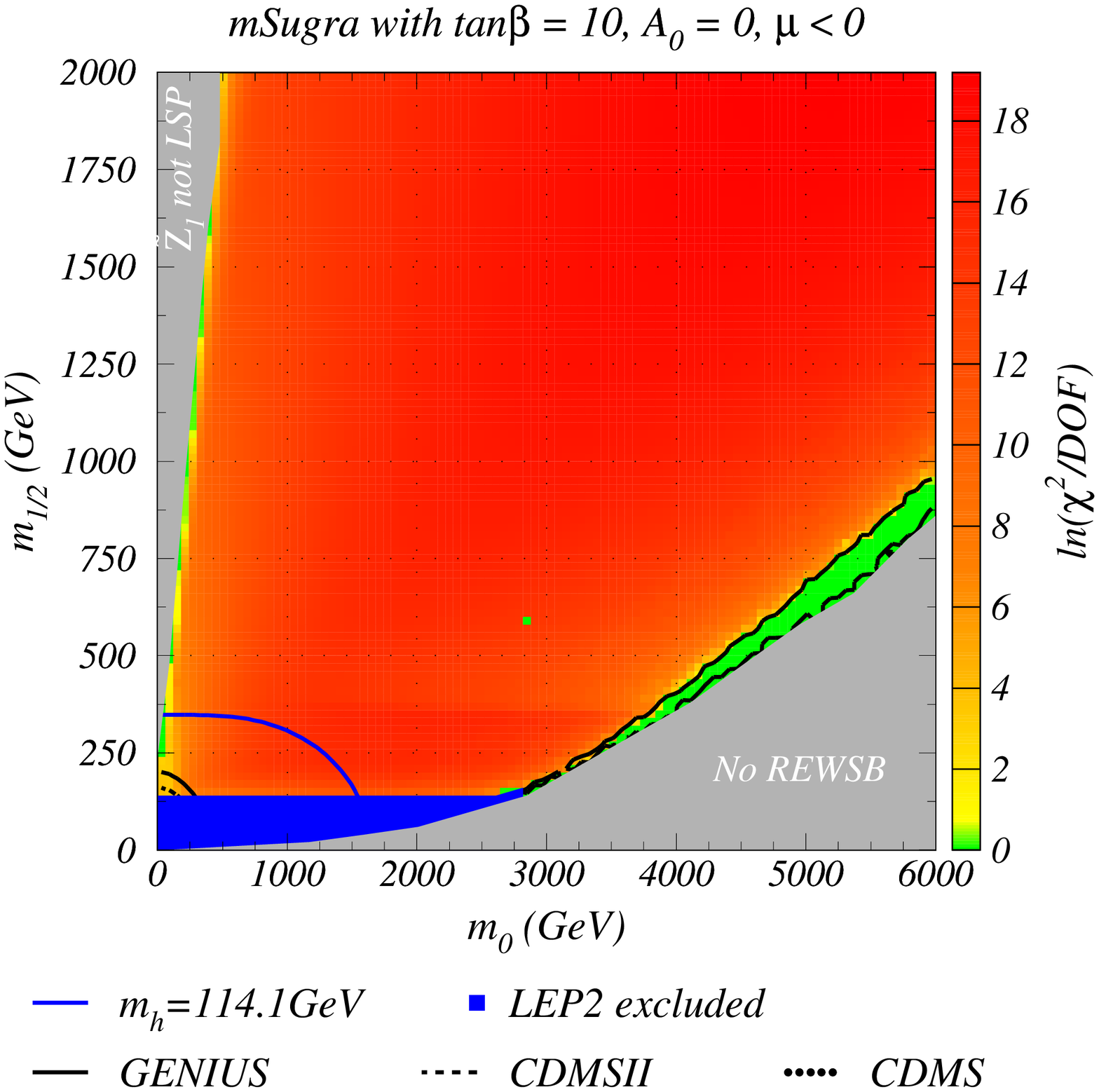,width=8cm} 
\epsfig{file=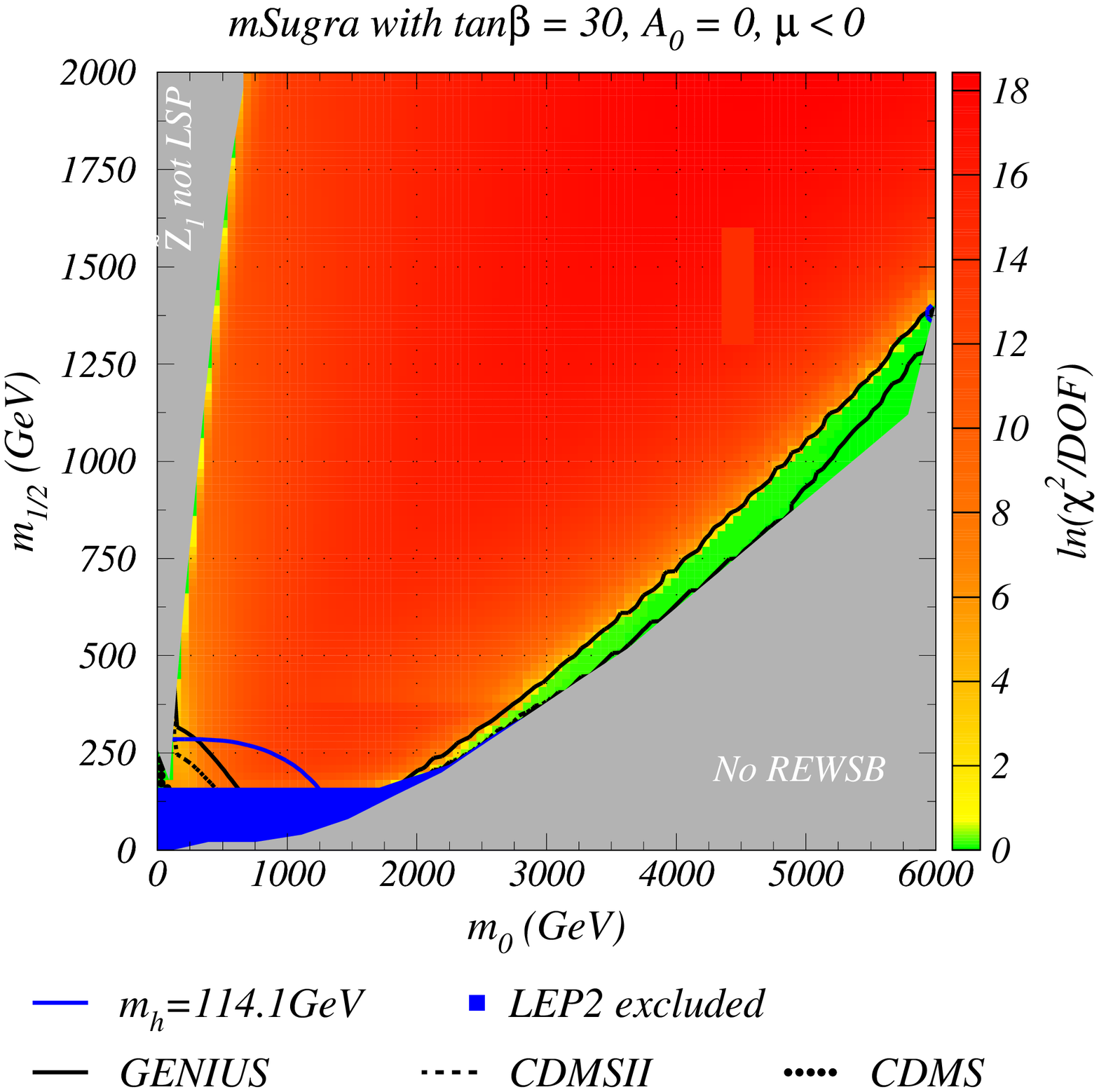,width=8cm}\\
\epsfig{file=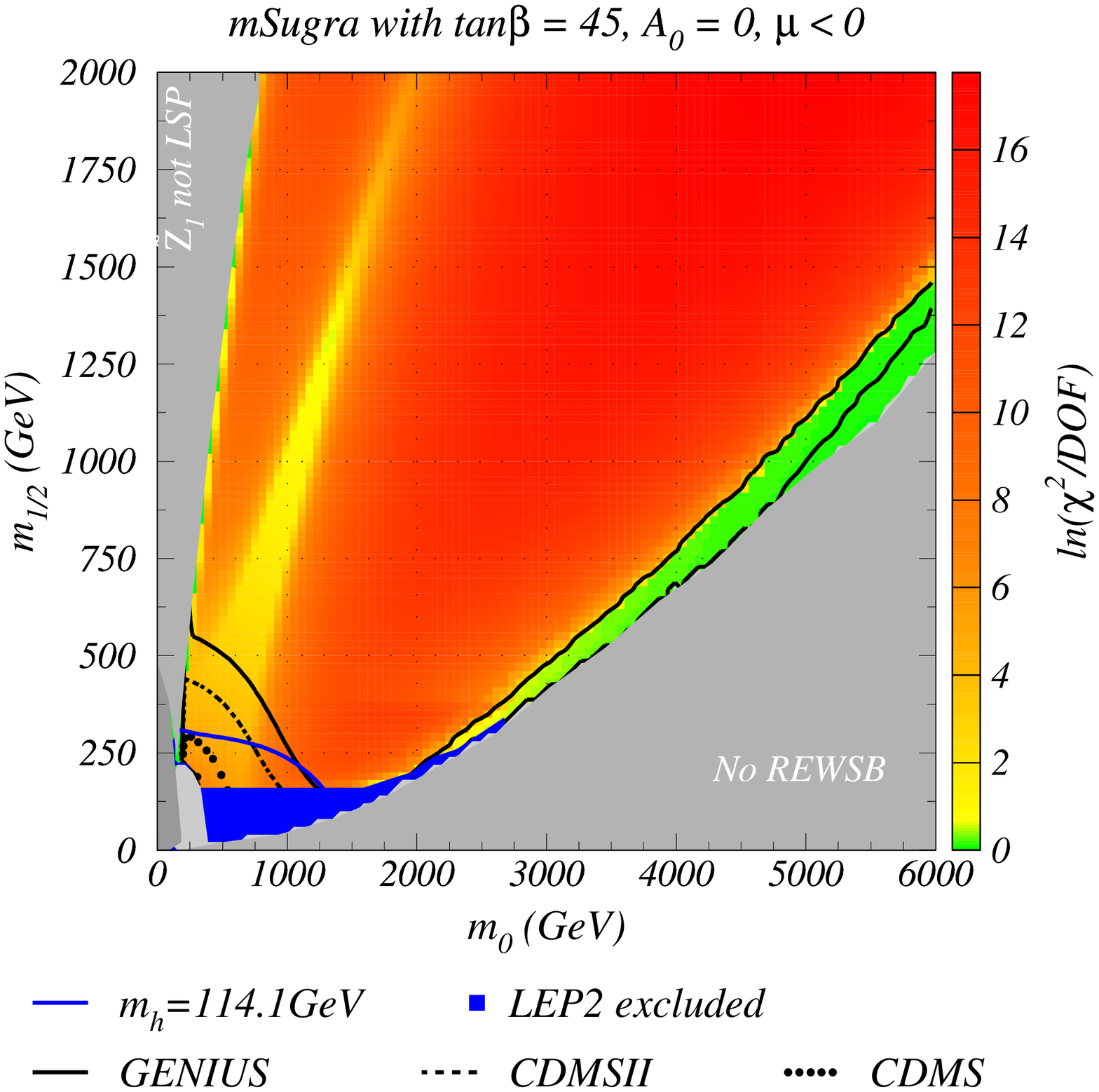,width=8cm} 
\epsfig{file=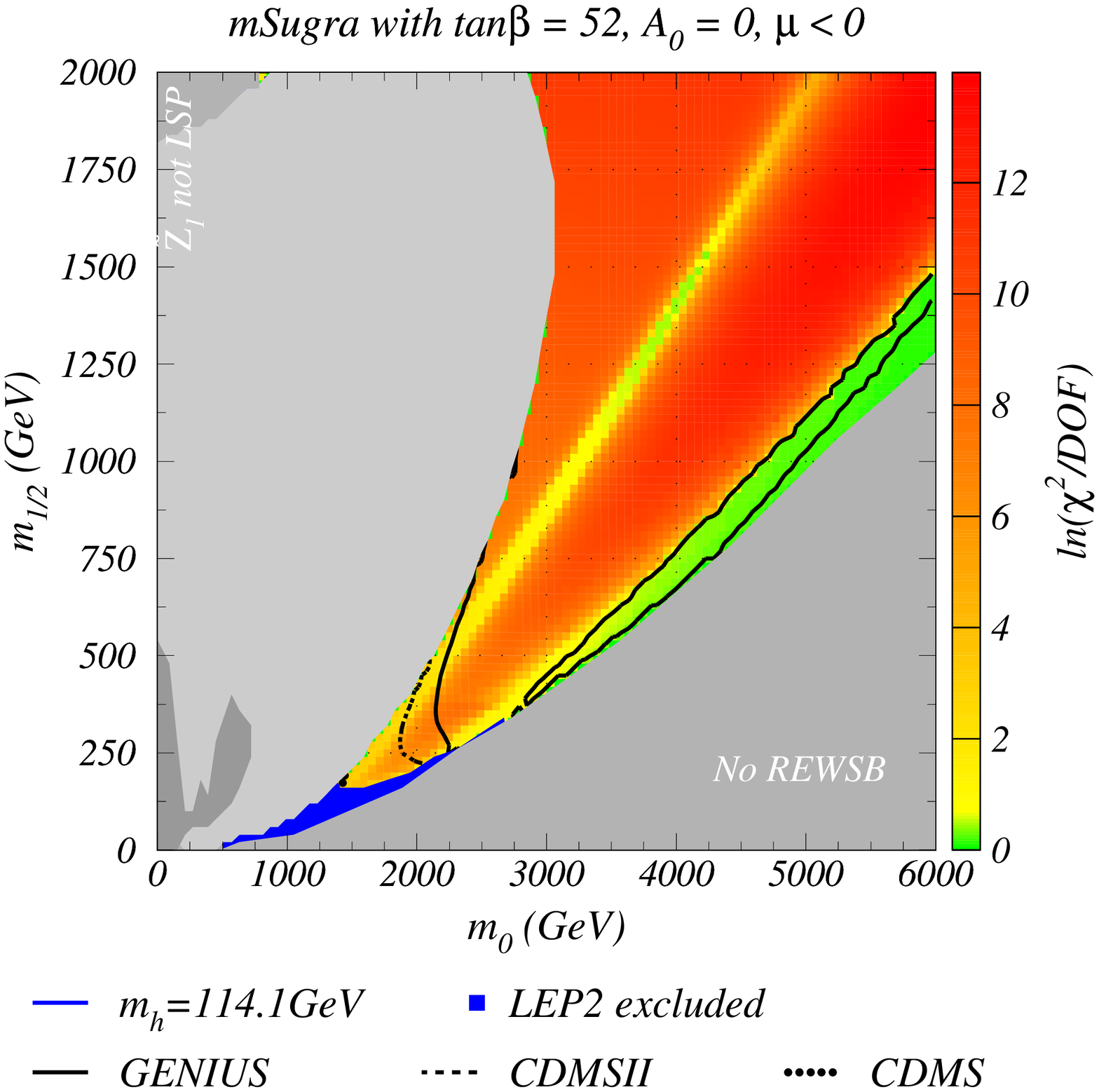,width=8cm}
\caption{Plot of $\chi^2 /dof$ for the mSUGRA model in the 
$m_0\ vs.\ m_{1/2}$ plane for $\mu <0$, $A_0=0$ and $\tan\beta =10$, $30$, 
$45$ and $52$.}
\label{fig:mun}
\end{figure}

In Fig. \ref{fig:mun}, we show the $m_0\ vs.\ m_{1/2}$ plane for
$\mu <0$ and values of $\tan\beta =$ {\it a.}) 10, {\it b.}) 30,
{\it c.}) 45 and {\it d.}) 52. 
The gray regions are excluded either by a non-neutralino LSP, or by a 
breakdown in the REWSB mechanism, signaled by $\mu^2 <0$ or by
$m_A^2<0$. The blue shaded region denotes points excluded by the 
LEP2 bound $m_{\tw_1}>103.5$ GeV, and the region below the blue 
contour is where $m_h<114.1$ GeV. 
We see from frame {\it a.}) that
only tiny green regions occur
along the excluded region at low $m_0$ where stau co-annihilation occurs,
or at the boundary of ``No REWSB'', where $|\mu |$ becomes small, 
the HB/FP region. The bulk region at low $m_0$ and 
$m_{1/2}$ is excluded by the LEP2 bound on $m_h$, but also has a large 
negative value of $a_\mu^{SUSY}$ and a large value
of $BF(b\to s\gamma )$. As $\tan\beta$ increases to 30, the HB/FP
region stands out as the main green region with low 
$\chi^2/dof$. Increasing $\tan\beta$ to 45 as in frame {\it c.}),
the HB/FP region remains the most viable, while a yellow
corridor of neutralino annihilation via $s$ channel $A$ and $H$
appears as splitting the plot. In addition, a tiny gray region at low
$m_0$ and $m_{1/2}$ has emerged, where $m_A^2 <0$.
As $\tan\beta$ increases to 52 as in frame {\it d.}), the $m_A^2<0$ constraint
has begun to overwhelm the plot at low $m_0$, pushing the Higgs annihilation
region to larger parameter values, where a few green points emerge. 
The HB/FP region remains robust. 
The parameter space becomes completely excluded at $\tan\beta$ 
values of 55 and higher.

The lesson from the $\mu <0$ plots is that the most robust region of
mSUGRA model parameter space is the HB/FP region, where the neutralino
has a significant higgsino component, so that efficient annihilation
(and co-annihilation) of neutralinos can occur in the early universe, 
in spite of quite heavy,
multi-TeV values of scalar masses. In fact, these scalar masses 
are sufficiently heavy to suppress possible SUSY $CP$ and flavor 
violating processes, while maintaining naturalness\cite{focus}.
They thus provide at least a partial solution to the SUSY flavor and 
$CP$ problems.

If in fact the relic cold dark matter is made of HB/FP
neutralinos, can these DM particles be detected by direct search
experiments? We show the reach of several direct
detection experiments (CDMS, CDMS2\cite{cdms}/CRESST\cite{cresst} 
and Genius\cite{genius}) 
for SUSY CDM by the
black contours, via the spin-independent neutralino-proton 
scattering rates as calculated in Ref. \cite{bbbo}.\footnote{
These contours emerge from digitizing the $\sigma\ vs.\ m_{\tz_1}$
reach contours presented by the various experimental groups. We have not
scaled the reach contours according to the value of the
neutralino relic density.}
Similar results
are given in Feng, Matchev and Wilczek\cite{fmw}, 
although their HB/FP region occurs at lower values of $m_0$
than ours.\footnote{Our HB/FP region as determined by ISAJET
is in good agreement with similar calculations by the 
programs Suspect, SoftSUSY and Spheno\cite{kraml}.}
It is gratifying to note that the most favored regions of 
parameter space are also accessible to direct search experiments, especially
large scale experiments such as Genius and Zeplin 4\cite{zeplin4}.

\begin{figure}
\epsfig{file=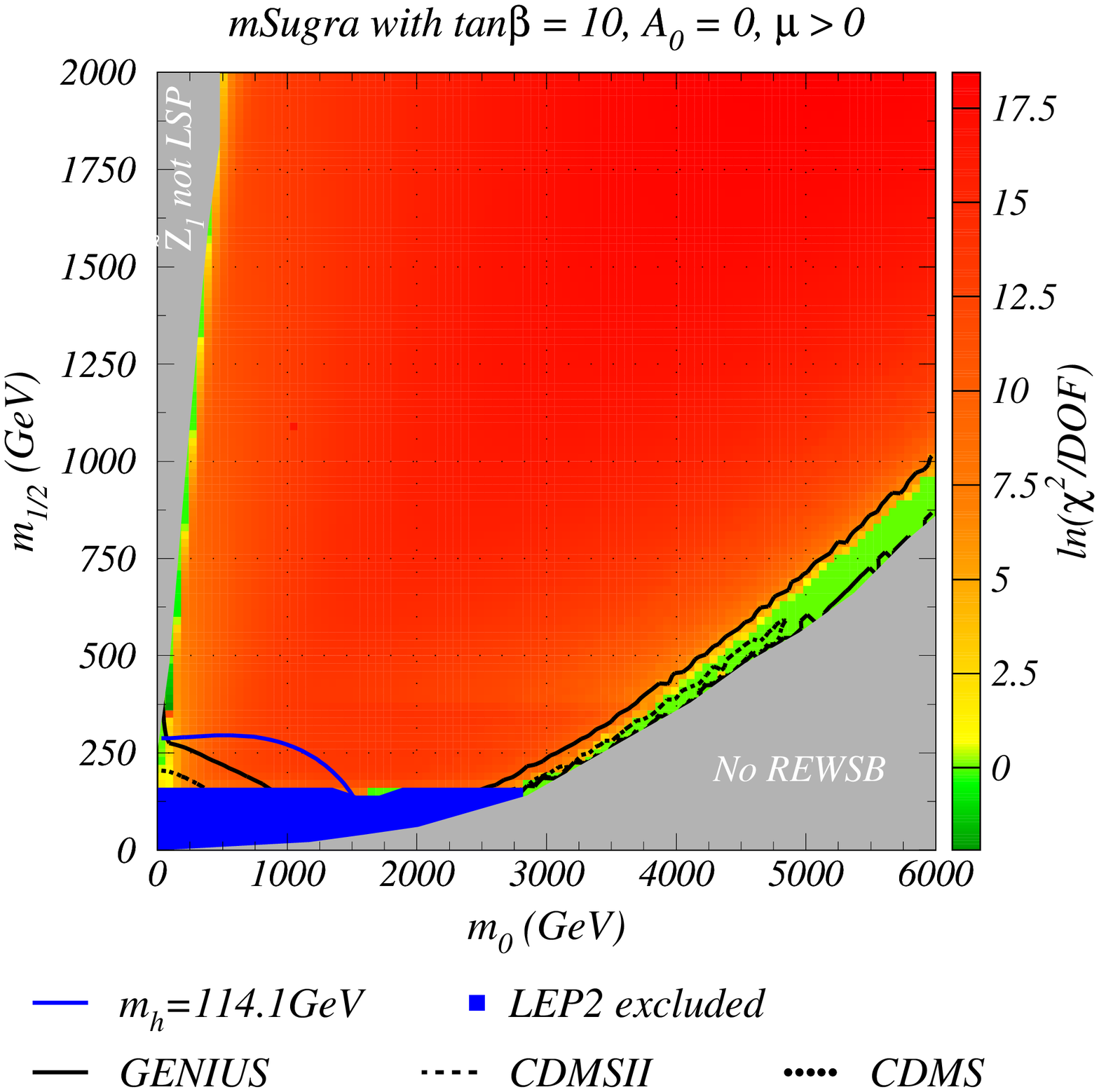,width=8cm} 
\epsfig{file=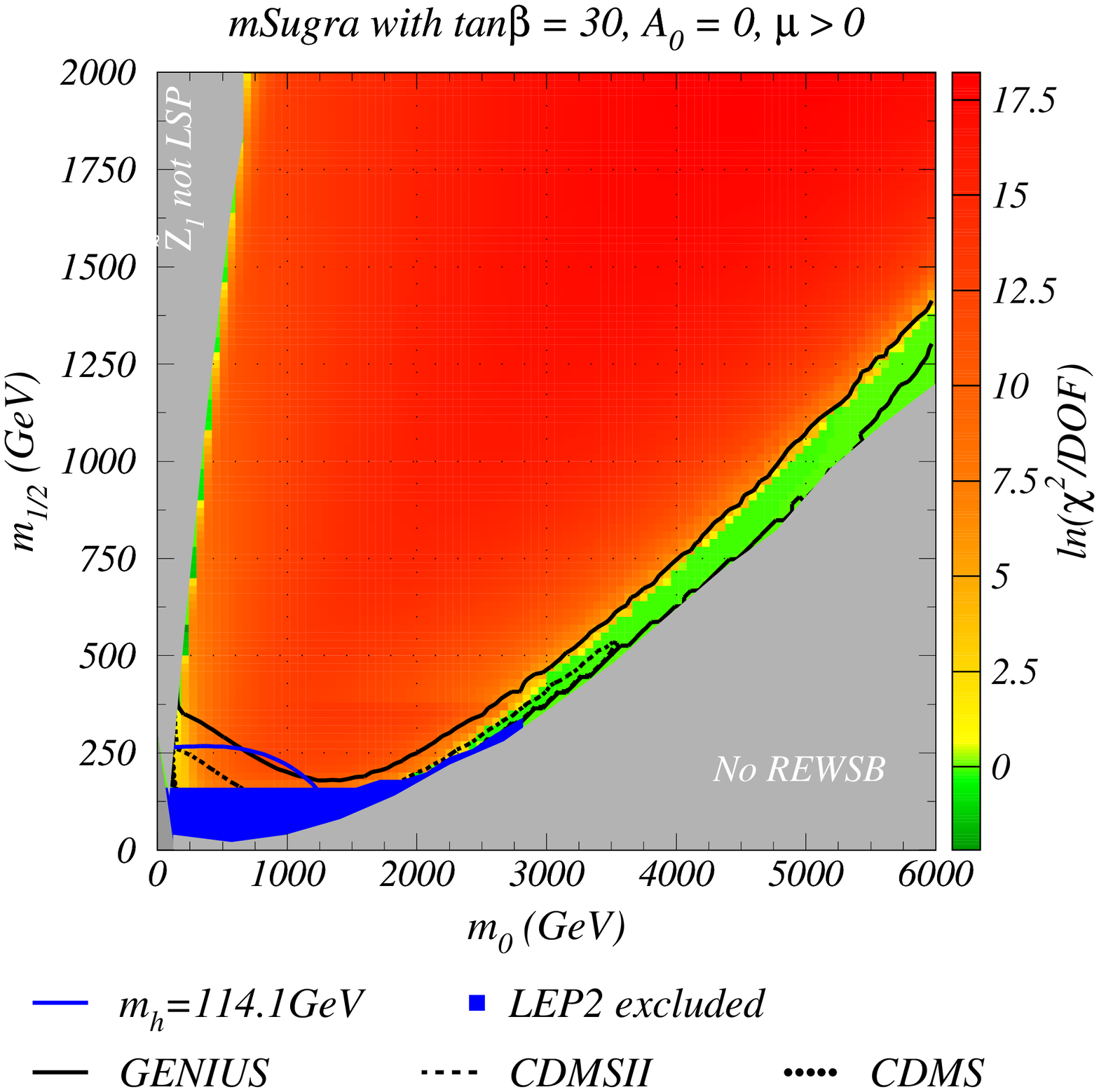,width=8cm}\\
\epsfig{file=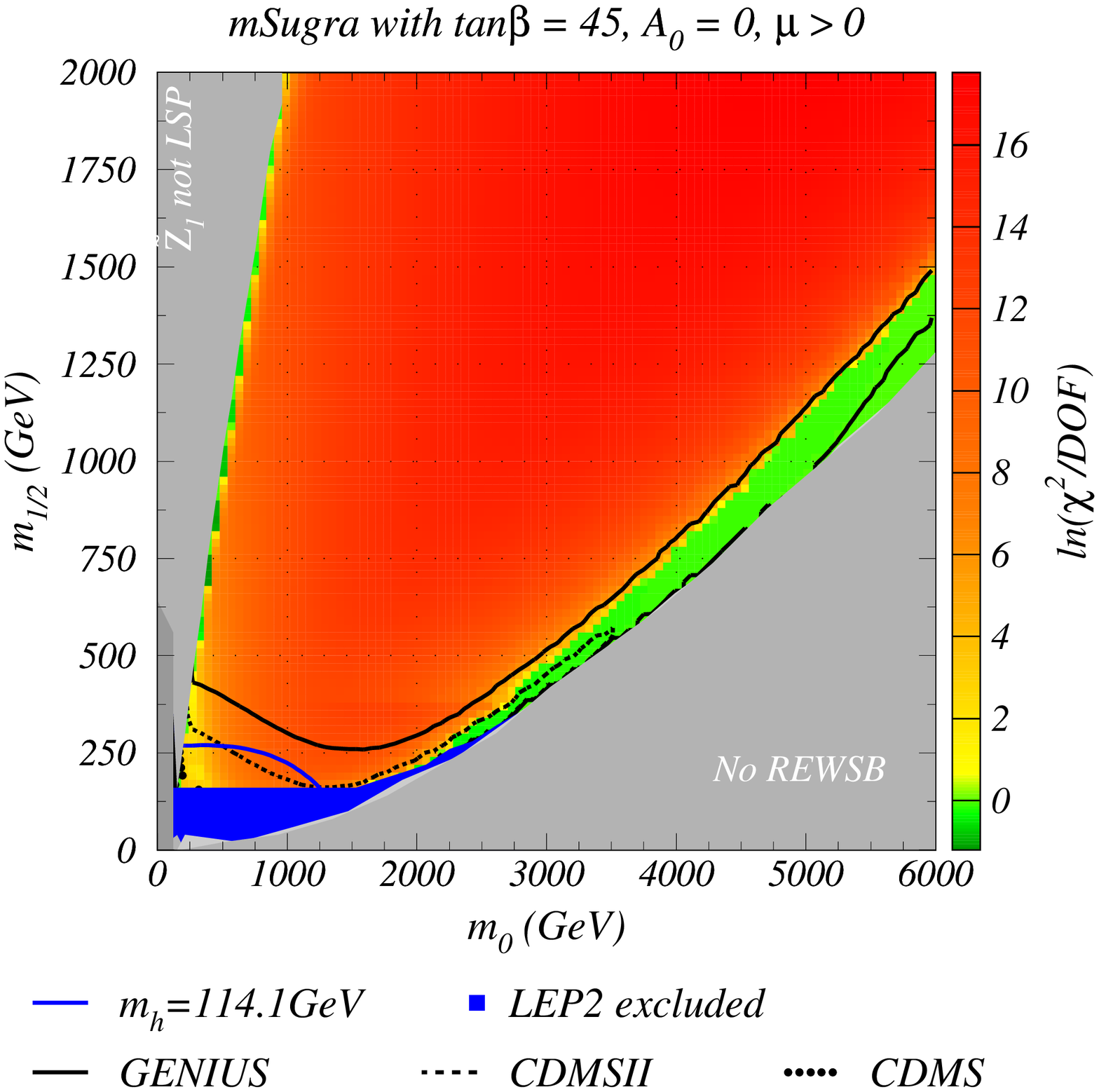,width=8cm} 
\epsfig{file=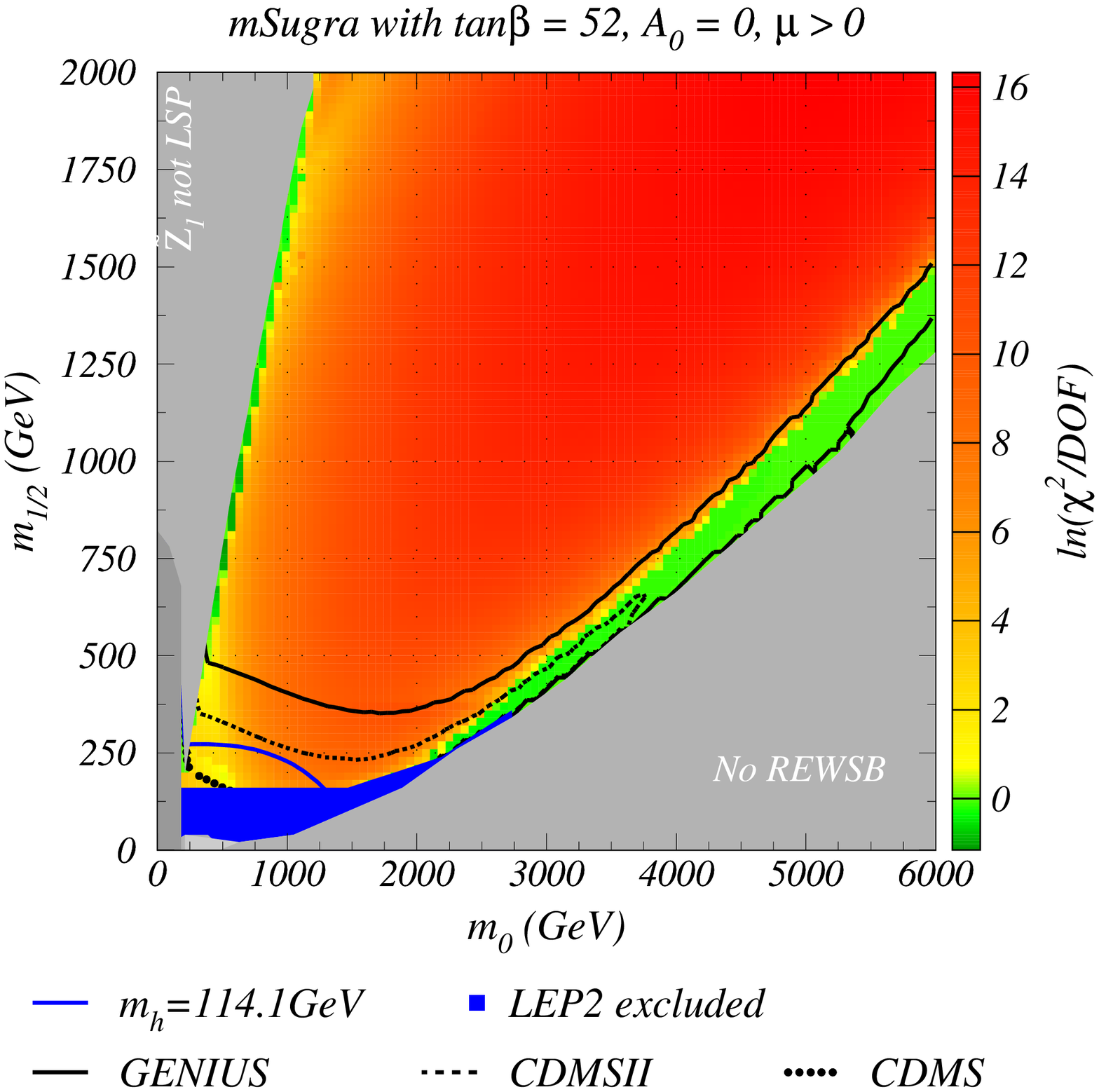,width=8cm}
\caption{Plot of $\chi^2 /dof$ for the mSUGRA model in the 
$m_0\ vs.\ m_{1/2}$ plane for $\mu >0$, $A_0=0$ and $\tan\beta =10$, $30$, 
$45$ and $52$.}
\label{fig:mup}
\end{figure}

In Fig. \ref{fig:mup}, we show the same mSUGRA model plane plots,
except for $\mu >0$. In this case, the values of 
$BF(b\to s\gamma )$ and especially $a_\mu^{SUSY}$ are more easily 
accommodated by the data\cite{sugcon}. The frame {\it a.}) for
$\tan\beta =10$ is qualitatively similar to the $\mu <0$ case, 
with the most favorable region again being the HB/FP region. As 
$\tan\beta $ increases to 30, the HB/FP region becomes even more 
promising. In addition, there are some tiny regions along the
stau co-annihilation border where a low $\chi^2/dof $ can be found.
The HB/FP region remains most promising for $\tan\beta$ values of
45 and 52. As in the $\mu <0$ case, it should be possible to
directly search for the HB/FP dark matter with CDMS2 and especially with
Genius and/or Zeplin4.

\begin{figure}
\epsfig{file=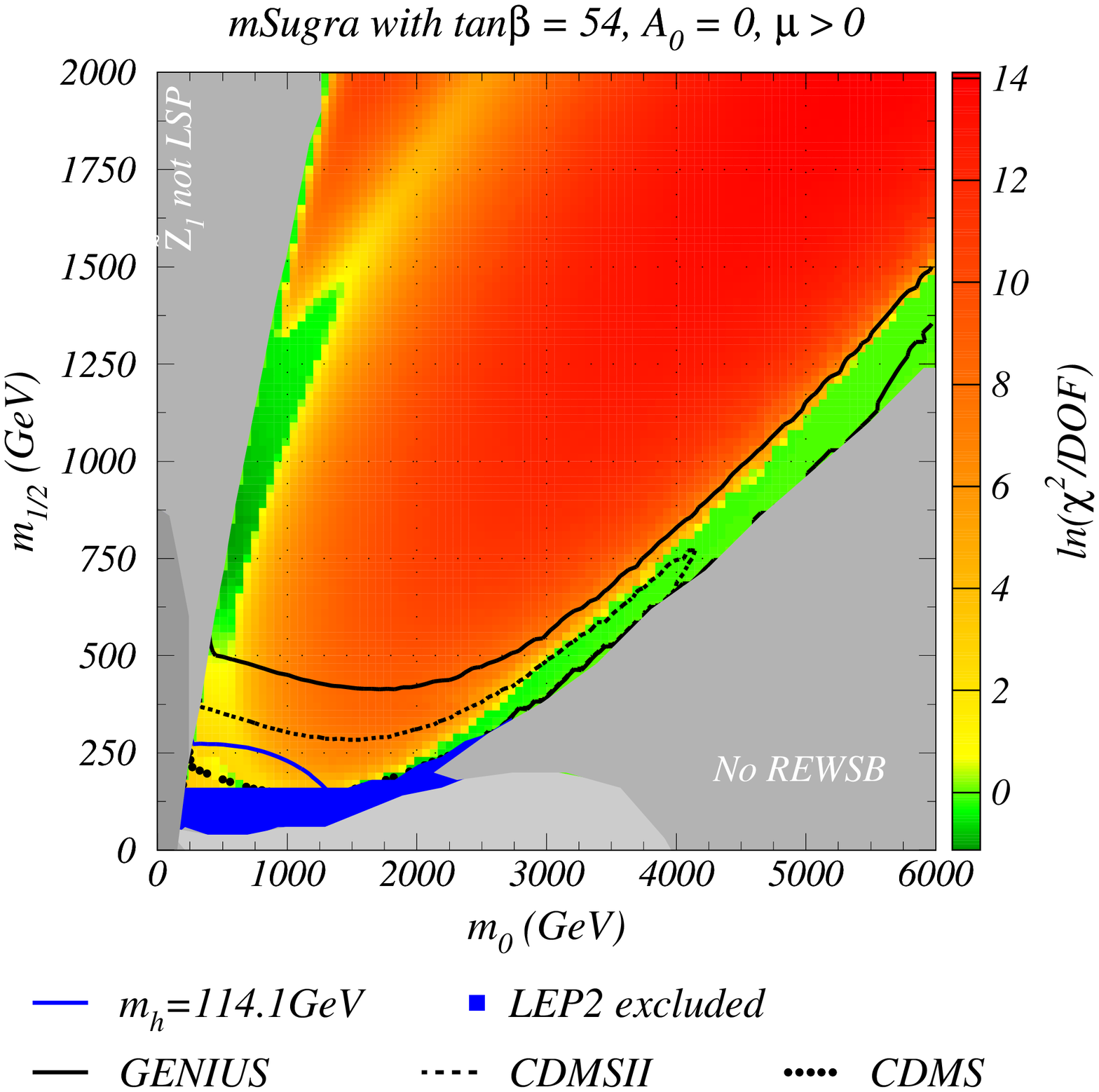,width=8cm} 
\epsfig{file=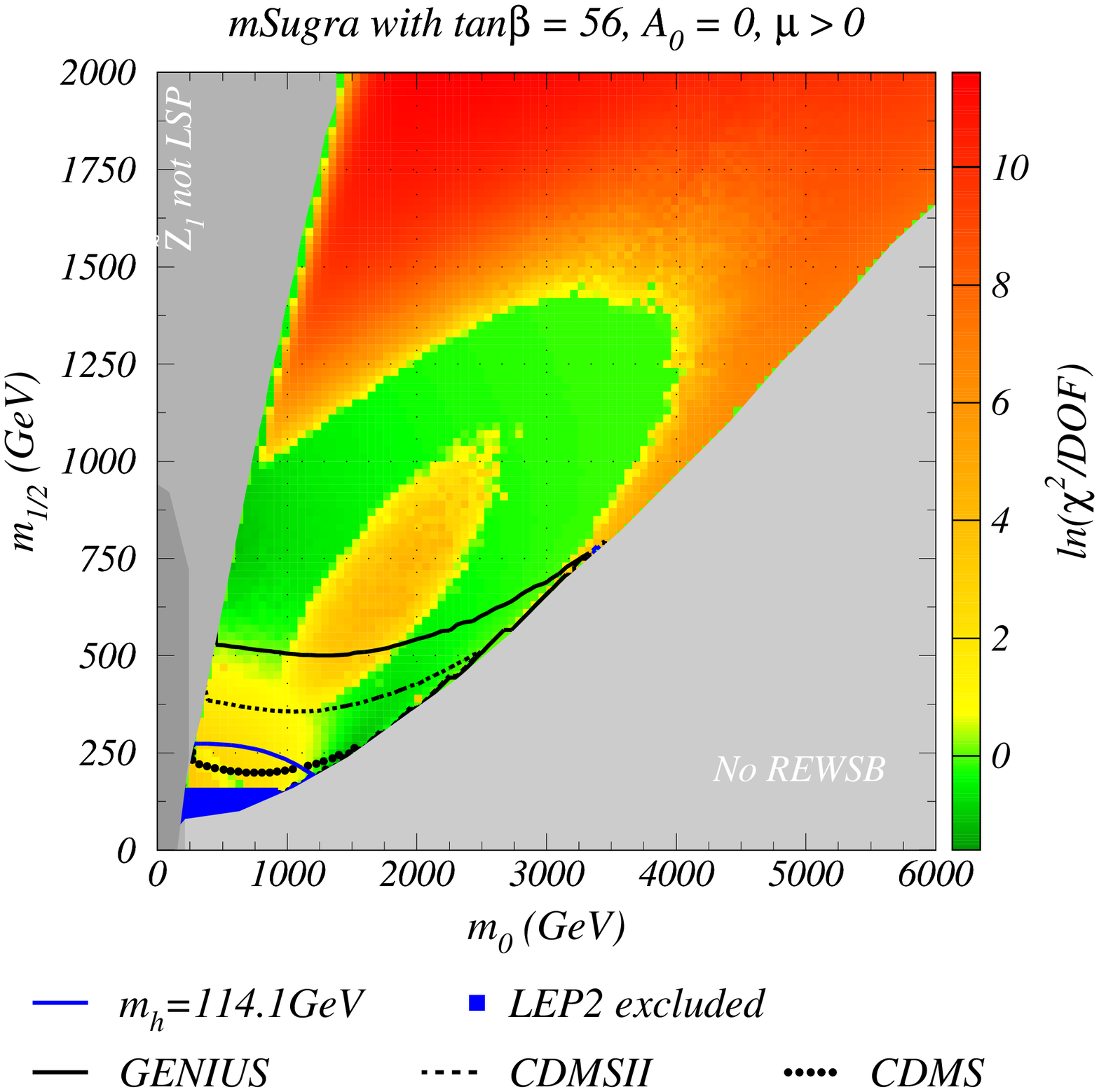,width=8cm}\\
\epsfig{file=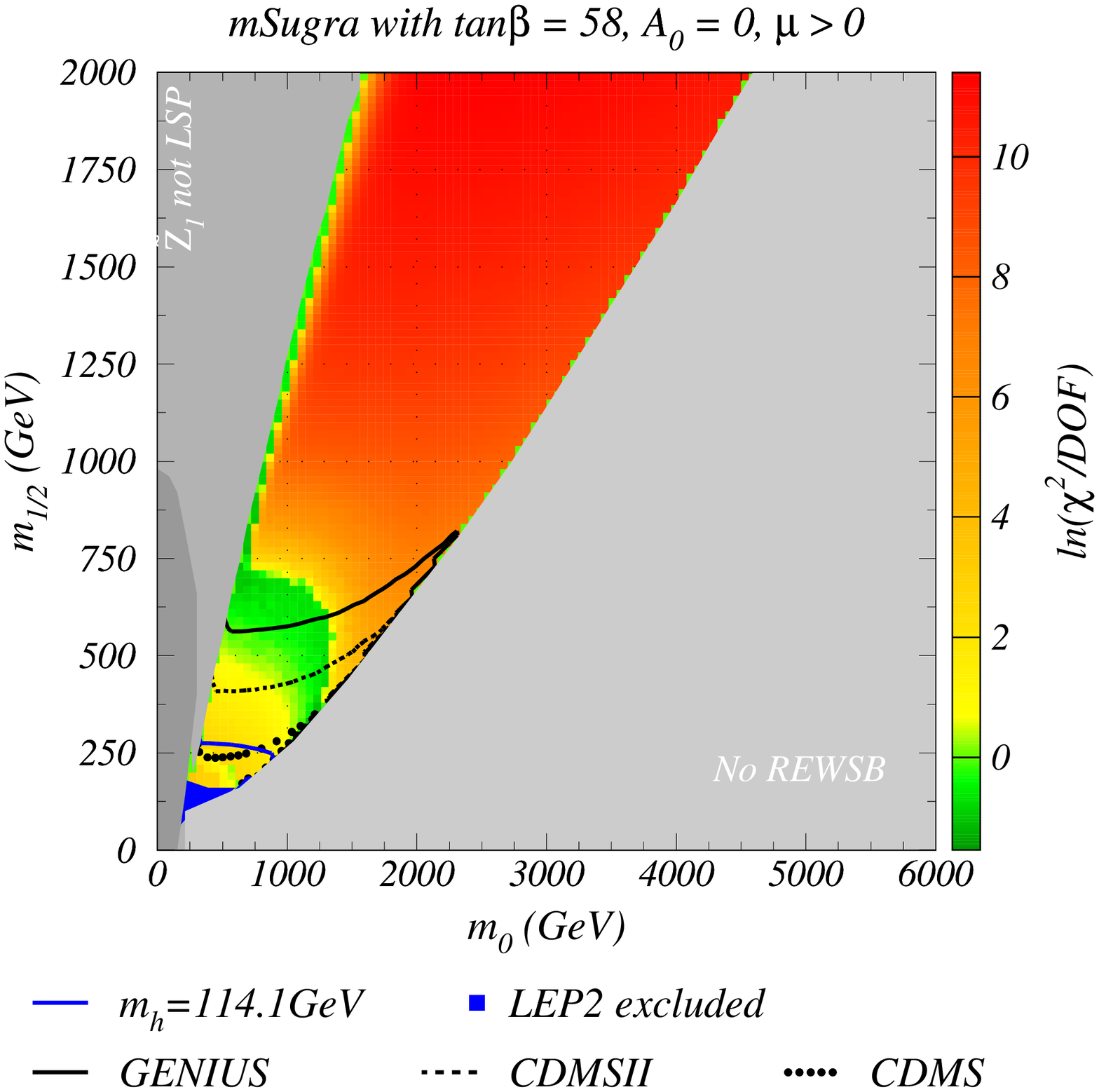,width=8cm} 
\epsfig{file=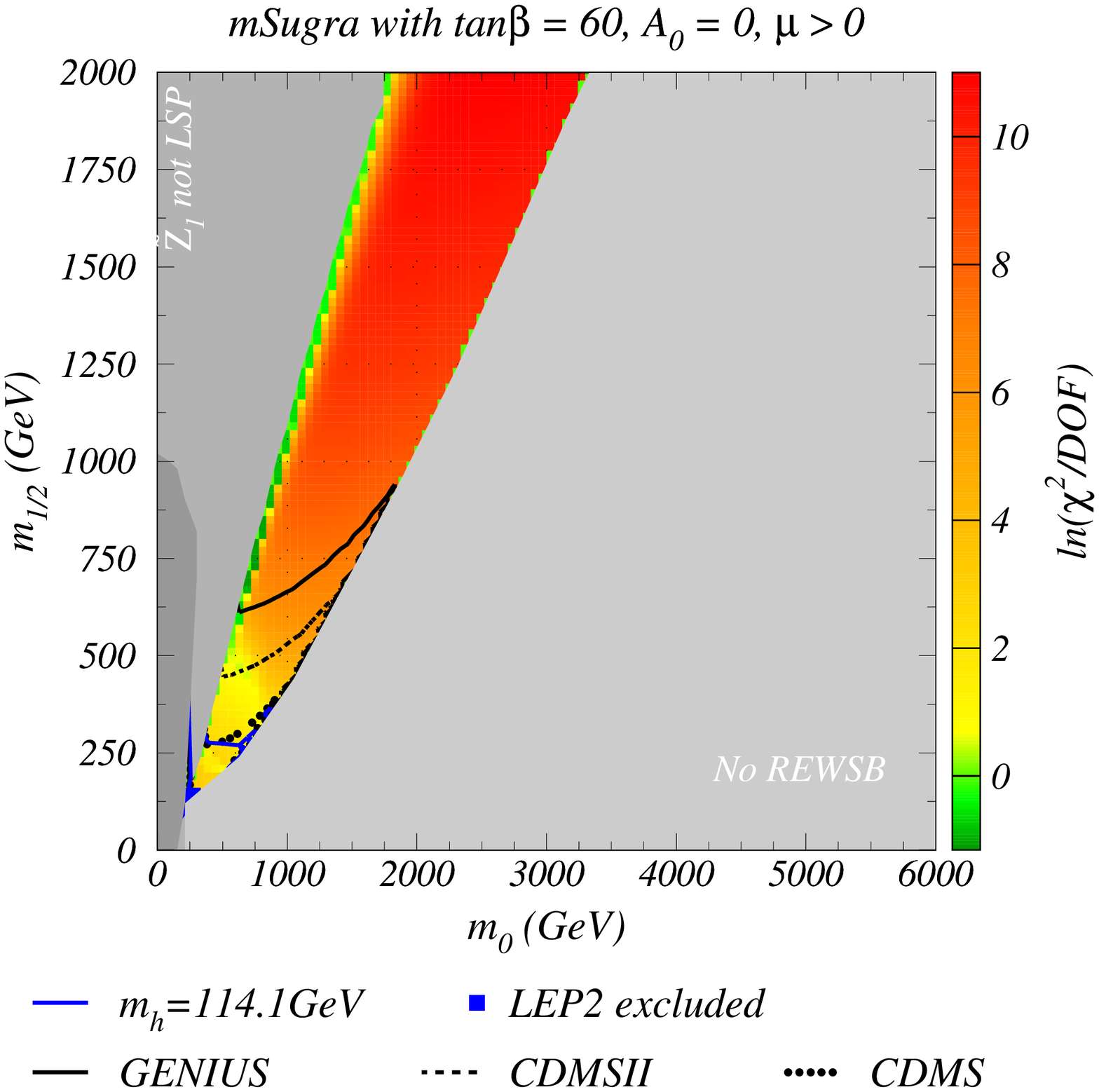,width=8cm}
\caption{Plot of $\chi^2 /dof$ for the mSUGRA model in the 
$m_0\ vs.\ m_{1/2}$ plane for $\mu >0$, $A_0=0$ and $\tan\beta =54$, $56$, 
$58$ and $60$.}
\label{fig:ltanb}
\end{figure}

Finally, we show in Fig. \ref{fig:ltanb} $\mu >0$ planes for
very large values of $\tan\beta =54$, 56, 58 and 60.\footnote{Such
large values of $\tan\beta$ fulfill naturalness conditions if
one loop corrections are included in evaluating the scalar 
potential\cite{bft}.}
In frame {\it a.}) for $\tan\beta =54$, the HB/FP region is even more
pronounced than in Fig. \ref{fig:mup}. 
Meanwhile, a corridor of $s$-channel Higgs annihilation is 
opening up at lower values of $m_0$, as shown by the green and yellow region.
For $\tan\beta =56$ in frame {\it b.}), the $m_A^2<0$ constraint
has begun to usurp the HB/FP region. In this case, now a broad
region of rapid neutralino annihilation has opened up where
$2m_{\tz_1}\sim m_{A,H}$. The region intermediate between these which 
is shaded yellow has $\Omega_{\tz_1}h^2$ just beyond the WMAP constraint.
The low $m_0$ and $m_{1/2}$ region has a somewhat higher $\chi^2$
value due to the value of $BF(b\to s\gamma )$ dropping below
$2\times 10^{-4}$.
As $\tan\beta$ increases to 58 in frame {\it c.}), a significant region
of resonance annihilation is evident. Much of it is accessible to direct
dark matter search experiments. Finally, in frame {\it d.}), only a fraction
of parameter space remains viable, but none of it with a low $\chi^2/dof$
value. The entire $m_0\ vs.\ m_{1/2}$ plane is excluded for 
$\tan\beta \ge 62$.

To summarize, we have combined the constraints from WMAP on the
neutralino relic density with constraints from $BF(b\to s\gamma )$ and
$(g-2)_\mu$ in a $\chi^2$ analysis which determines favorable and
unfavorable regions of mSUGRA model parameter space. We find the bulk
neutralino annihilation region (A.) at low $m_0$ and $m_{1/2}$
essentially ruled out by constraints from LEP2, $BF(b\to s\gamma )$ and
$(g-2)_\mu$. In addition, the stau co-annihilation region (B.)
has only tiny favorable regions, which would require fine-tuning
of parameters to satisfy all constraints. The HB/FP region (D.)
emerges as a significant region satisfying all constraints over a
wide range of $\tan\beta$ values, and also offers at least a partial
solution to the SUSY flavor and $CP$ problems. In addition, 
the neutralino resonance annihilation regions (C.) for $\mu >0$ and
large $\tan\beta$ can satisfy all constraints. 
The favorable HB/FP regions are all accessible to direct dark
matter search experiments, and much of it should be accessible to TeV scale
linear colliders\cite{nlc}, since $|\mu |$ is small, and the lightest chargino
frequently lighter than $\sim 500$ GeV. The HB/FP region should also
be accessible at LHC searches as long as $m_{1/2}$ is not too large,
so that gluino pair production occurs at a high enough rate\cite{lhc}. 
 
\section*{Acknowledgments}
 
This research was supported in part by the U.S. Department of Energy
under contract number DE-FG02-97ER41022.
	
%

\end{document}